\documentclass[10pt, twocolumn]{article}
\usepackage[margin=1in]{geometry}
\usepackage{amsmath}
\usepackage{abstract}
\usepackage{amsfonts,color}
\usepackage{amssymb}
\usepackage{graphicx}
\usepackage{hyperref}
\begin{document}
\title{\bf{Bifurcations of a Van der Pol oscillator in a double well}}
\author
{ Satadal Datta\\
Harish-Chandra research Institute, HBNI, Chhatnag Road, Jhunsi, Allahabad-211019, INDIA\\
\date{}
satadaldatta1@gmail.com}
\twocolumn[
\maketitle
\begin{onecolabstract}
We study the changes in the phase portrait of a Van der Pol oscillator in a double well with the change in the shape of the well. Both the symmetric and asymmetric double wells are considered. We find that the shape of the well plays a very important role in the dynamics of the oscillator such as with the change in shape of the well, the stability of the fixed points of the system changes as well as limit cycles appear and are being destroyed.
\end{onecolabstract}
]
\vspace*{0.33cm}
\section{Introduction}
Self sustained oscillatory systems were first found and studied in the late nineteenth century\cite{a}. This kind of oscillations are reflected in the phase portrait as isolated periodic orbits, i.e; limit cycles. French mathematician Henri Poincar$\rm\hat{e}$ first introduced the concept of limit cycle in his theory of differential equations. Dutch radio physicist and electrical engineer Balthasar van der Pol introduced his famous Van der Pol equation in studying spontaneous self-sustained oscillating current in a triode\cite{c}; known as relaxation oscillations. The famous non linear Van der Pol equation is given by
\begin{equation*}
\ddot{x}+k\dot{x}(x^2-1)+x=0
\end{equation*}
Self sustained oscillatory systems, i.e; which have isolated periodic orbit, namely limit cycle, are ubiquitous in nature and they are studied broadly in several occasions; for example, in physiology\cite{d}\cite{e}\cite{f}, in studying circadian cycle of plants\cite{g}\cite{h} etc.\\
In this paper, we study the bifurcations of a double well Van der Pol oscillator. We find that with the change in the shape of the well the phase portrait of the system undergoes a series of dramatic events like Hopf bifurcations, creation of bigger limit cycle by merging of two limit cycles of same kind (stable-stable or unstable-unstable), annihilation of limit cycles by collision of two limit cycles of opposite kind (stable-unstable) etc.\\\\\\\\
\section{Symmetric double well Van der Pol oscillator}
We begin with a Van der Pol oscillator in a symmetric double well.
\begin{equation}
\ddot{x}+k\dot{x}(x^2-1)+(\lambda x^3-x)=0
\end{equation} 
$k$ is damping coefficient can be positive or negative real number, $k\neq 0$, $\lambda>0$.\footnote{In the last term of equation (1), one can work with $(ax^3-bx)$ with $a,~b>0$ instead of $(\lambda x^3-x)$; that would be just a constant multiplication of the form we have taken, hence the main characteristics of the potential function wouldn't differ.}
Hence the double well potential is $(\frac{1}{4}\lambda x^4-\frac{1}{2}x^2)$ with zeros at $x=0,~x=\pm\sqrt{\frac{2}{\lambda}}$, with a maxima at $x=0$ and with two minimas, i.e; the wells, at $x=\pm\sqrt{\frac{1}{\lambda}}$ . 
The above dynamical system can be written as 
\begin{equation*}
\dot{x}=y
\end{equation*}
\begin{equation}
\dot{y}=-ky(x^2-1)-(\lambda x^3-x)
\end{equation} 
The fixed points are $(0,~0),~(\pm\sqrt{\frac{1}{\lambda}},~0)$. Varying $x,~y$ linearly around the fixed point $(0,~0)$ gives
\begin{equation}
\begin{pmatrix} \delta \dot{x} \\ \delta \dot{y} \end{pmatrix}=
\begin{pmatrix} 0 & 1\\ 1 & k \end{pmatrix}
\begin{pmatrix} \delta x \\ \delta y \end{pmatrix} =
\hat{A} \begin{pmatrix} \delta x \\ \delta y \end{pmatrix}
\end{equation}
Trace of the matrix, $\hat{A}=Tr(\hat{A})=k\neq 0$ and determinant of the matrix, $det(\hat{A})=-1$, hence the fixed point is a hyperbolic fixed point\cite{i}. Using Hartman-Grobman theorem\cite{j}-\cite{l}, we can state that the topology of the phase portrait near the origin is of saddle type. Again varying $x$ and $y$ near fixed points $(\pm\sqrt{\frac{1}{\lambda}},~0)$, we get
\begin{equation}
\begin{pmatrix} \delta \dot{x} \\ \delta \dot{y} \end{pmatrix}=
\begin{pmatrix} 0 & 1\\ -2 & -k(\frac{1}{\lambda}-1) \end{pmatrix}
\begin{pmatrix} \delta x \\ \delta y \end{pmatrix} =
\hat{B} \begin{pmatrix} \delta x \\ \delta y \end{pmatrix}
\end{equation}
$Tr(\hat{B})= -k(\frac{1}{\lambda}-1)$, $det(\hat{B})=2$. Hence for small $k$, each of fixed points $(\pm\sqrt{\frac{1}{\lambda}},~0)$ is locally unstable spiral for $\lambda>1,~k>0$ and locally stable spiral for $\lambda<1,~k>0$. For $k<0$, each of the fixed points $(\pm\sqrt{\frac{1}{\lambda}},~0)$ is locally stable spiral for $\lambda>1$ and locally stable spiral for $\lambda<1$, assuming $k$ to be small. We find numerically that for $\lambda>1$, $\exists$ a stable limit cycle for $k>0$ and an unstable limit cycle for $k<0$. Hence at $\lambda=1$ there is a Hopf bifurcation\cite{m}\cite{n} happening. Using Poincar$\rm\hat{e}$-Bendixson theorem \cite{p}, for $k>0$, the Hopf bifurcation is sub critical, i.e; as $\lambda$ becomes less than $1$ two unstable limit cycles appear around each of $(\pm\sqrt{\frac{1}{\lambda}},~0)$ and for $k<0$, the Hopf bifurcation is super critical, i.e; as $\lambda$ becomes less than $1$ two stable limit cycles appear around each of $(\pm\sqrt{\frac{1}{\lambda}},~0)$. Therefore, as $\lambda$ becomes just less than $1$, the phase portrait of the system contains three limit cycles; around each of $(\pm\sqrt{\frac{1}{\lambda}},~0)$, there are two limit cycles of same kind and there is another existing bigger boundary limit cycle of opposite kind illustrated in figure 2. We calculate analytically the amplitude of each limit cycle surrounding $(\pm\sqrt{\frac{1}{\lambda}},~0)$ created by the Hopf bifurcation for $\lambda\lesssim 1,~k\simeq 0$ by using Krylov-Bogolyubov approximation method. Displacement $x(t)$ at the neighbourhood of $x=\frac{1}{\sqrt{\lambda}}$, is written as
\begin{equation*}
x(t)=\frac{1}{\sqrt{\lambda}}+x'(t)
\end{equation*}
Putting this value of $x(t)$ in equation (1) we get 
\begin{equation}
\ddot{x'}+k\dot{x'}(x'^2+\frac{2}{\sqrt{\lambda}}x'+\frac{1}{\lambda}-1)+(2x'+3\sqrt{\lambda}x'^2+\lambda x'^3)=0
\end{equation}
As $x'$ is small, $k$ is small and as $\lambda=(1-\epsilon)$ where $\epsilon\gtrsim 0$, we approximate $x'$ as
\begin{equation}
x'(t)=A(t)cos(\omega t+\phi(t))
\end{equation}
where $\omega=\sqrt{2}$ because the linear variation of $(\lambda x^3-x)$ around $x=\frac{1}{\sqrt{\lambda}}$ gives a factor of $2$ which is also present in equation (5), i.e; the coefficient of $x'$, hence we find $\omega^2$. Basically, we seek for periodic solution, i.e; the limit cycle solution. Amplitude and phase $A(t)$ and $\phi(t)$ are slowly varying function of time $t$. Hence we neglect $\ddot{A}, ~\ddot{\phi}(t)$. We have 
\begin{equation*}
\dot{x'}=\dot{A}cos(\omega t+\phi(t))-A(\omega+\dot{\phi})sin(\omega t+\phi (t))
\end{equation*}
\begin{equation*}
\ddot{x'}=-2\dot{A}\omega sin(\omega t+\phi (t))-A(\omega^2+2\omega\dot{\phi})cos(\omega t+\phi(t))
\end{equation*}
Using equation (5) and using the orthogonality of sine and cosine functions, we get after finding the coefficient of $cos(\omega t+\phi(t))$ and $sin(\omega t+\phi (t))$;
\begin{eqnarray}
& &-2\omega A(t)\dot{\phi}(t)+k\dot{A}(t)\left(\frac{1}{\lambda}-1\right)+\frac{3}{2}k\dot{A}(t)A(t)^2\nonumber\\
& &+\frac{3\lambda}{4}A(t)^3=0
\end{eqnarray}
and 
\begin{eqnarray}
& &-2\dot{A}(t)\omega-kA(t)\left(\omega+\dot{\phi}(t)\right)\left(\frac{1}{\lambda}-1\right) \nonumber\\
& &+\frac{kA(t)^3}{4}\left(\omega+\dot{\phi}(t)\right)=0
\end{eqnarray}
From equation (8) we seek for fixed point solution of $A(t)$, i.e; we seek constant amplitude solution. Hence putting $\dot{A}(t)$ to be zero, we get
\begin{equation}
A=0~or,~ A=2\sqrt{\frac{1}{\lambda}-1}\simeq 2\sqrt{\epsilon}
\end{equation} 
Equation (7) gives
\begin{equation}
\dot{\phi}=\frac{3\lambda A^2}{8\sqrt{2}}\simeq\frac{3}{2\sqrt{2}}\epsilon
\end{equation}
Hence 
\begin{equation}
x(t)\simeq\frac{1}{\sqrt{\lambda}}+2\sqrt{\epsilon}~cos\left(\left(\sqrt{2}+\frac{3}{2\sqrt{2}}\epsilon\right)t+\Phi\right)
\end{equation}
where $\Phi$ is the constant phase factor. Hence at $\lambda=1$, the periodic orbit is of zero radius.\\
From equation (9), it is also obvious that for $\lambda>1$ finite amplitude solution does not exist. At $\lambda=(1-\epsilon),~\epsilon>0$, the fixed point $(\frac{1}{\sqrt{\lambda}},0)$ is stable for $k>0$, hence using Poincar$\rm\hat{e}$-Bendixson theorem the limit cycle produced due to Hopf bifurcation is unstable and for $k<0$, the limit cycle is stable. One can also show this from equation (7), by varying $A(t)$ around $A=0$ and $A=2\sqrt{\frac{1}{\lambda}-1}$ and by dropping $k\dot{\phi}$ term by taking small $k$ limit. Thus we find the amplitude of the limit cycle at small $k$ limit at $\lambda\lesssim 1$. One can find the same results around $x=-\frac{1}{\sqrt{\lambda}}$ because of symmetry. \\
Hence for $\lambda >1$, there is only one limit cycle in the phase portrait of the system. 
%\newpage
\begin{figure}[hbtp]
%\centering
\includegraphics[height=8.7cm, width=6cm, angle=-90]{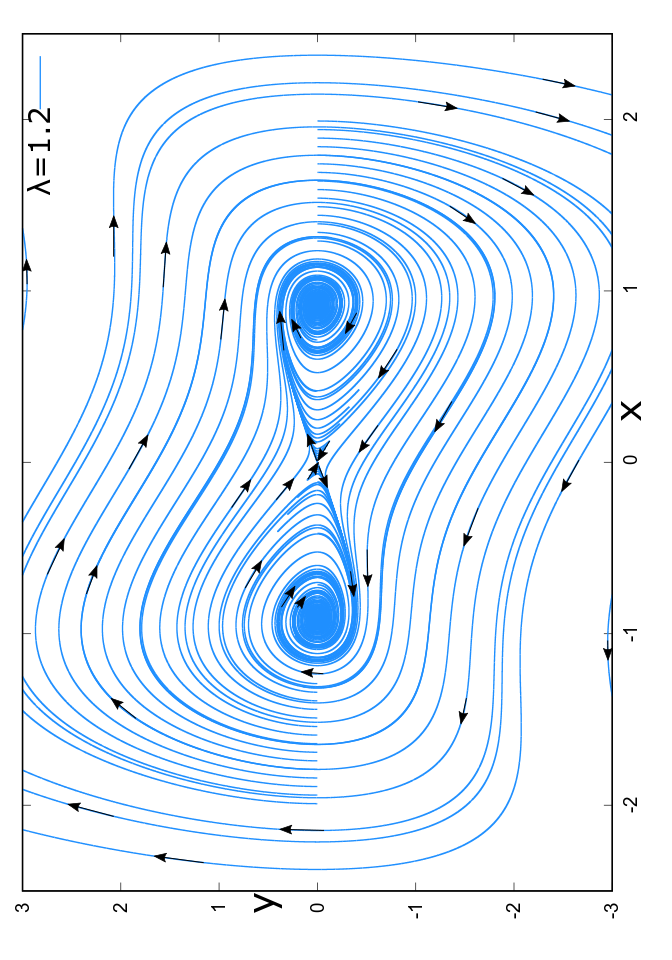}
\caption{$\lambda>1,~k=-1$; one unstable limit cycle and at $\lambda=1$, Hopf bifurcation occurs.}
\end{figure}
\begin{figure}[hbtp]
%\centering
\includegraphics[height=8.7cm, width=6cm, angle=-90]{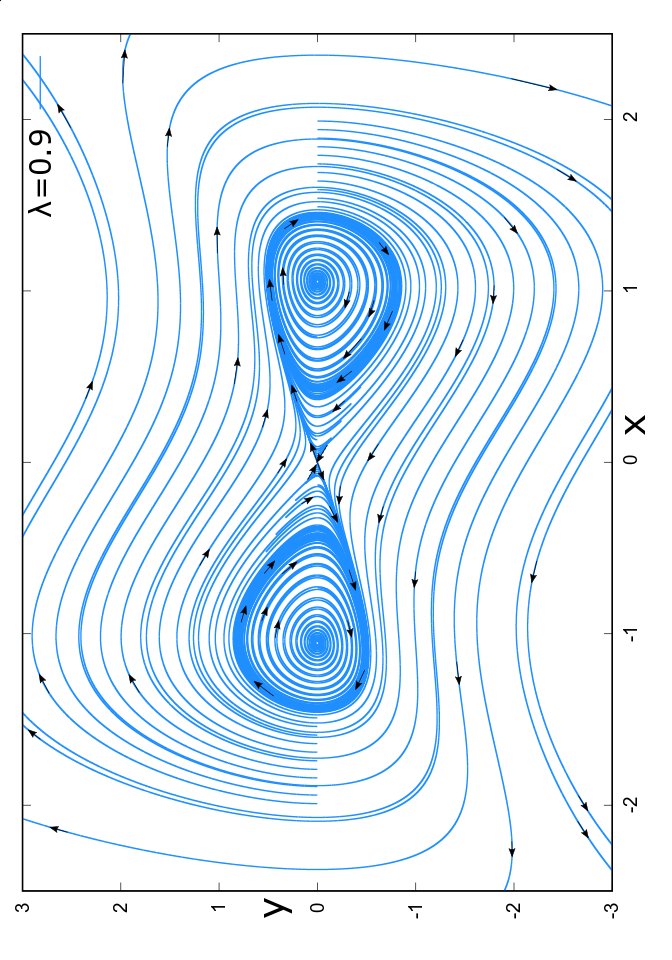}
\caption{$\lambda=0.9,~k=-1$; one unstable limit cycle and two newly formed stable limit cycles.}
\end{figure}\begin{figure}[hbtp]
%\centering
\includegraphics[height=8.7cm, width=6cm, angle=-90]{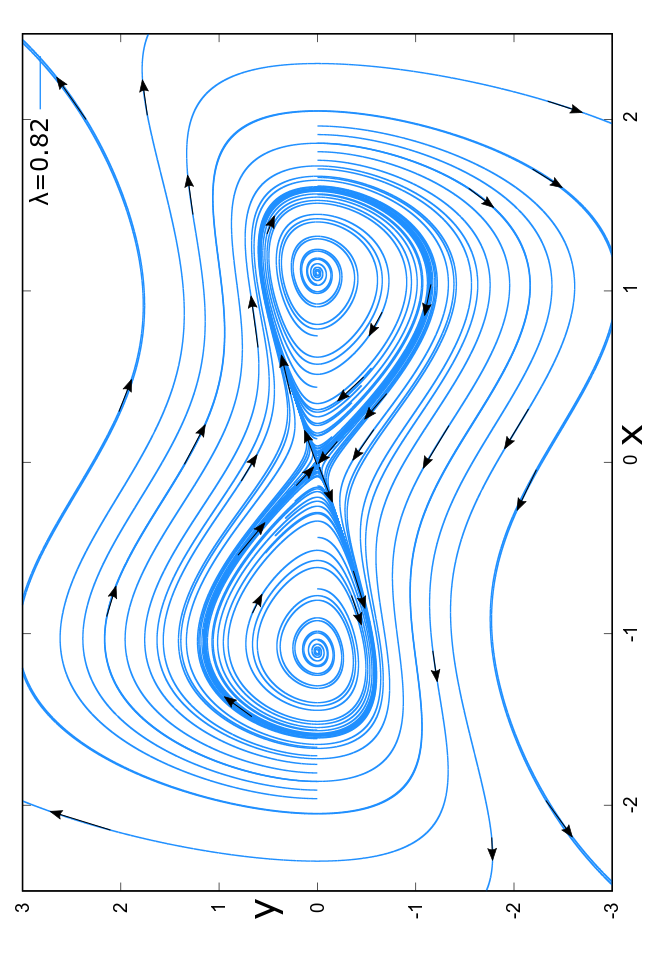}
\caption{$\lambda=0.82,k=-1$; two stable limit cycles touch each other, homoclinic orbit appears.}
\end{figure}
\begin{figure}[hbtp]
%\centering
\includegraphics[height=8.7cm, width=6cm, angle=-90]{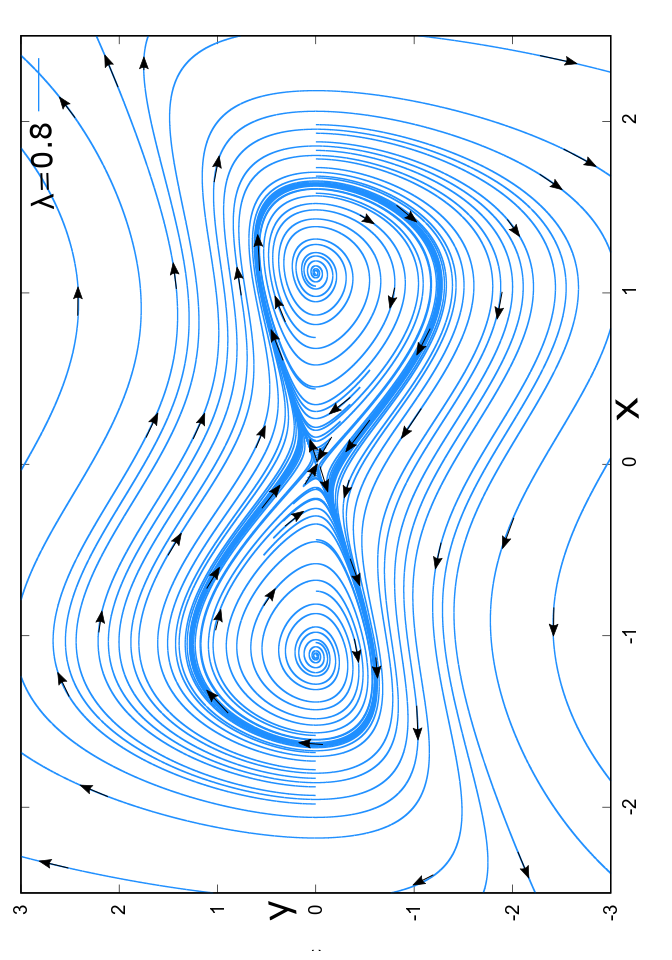}
\caption{$\lambda=0.8,~k=-1$, one stable limit cycle formed by merging and one unstable limit cycle.}
\end{figure}
\begin{figure}[hbtp]
%\centering
\includegraphics[height=8.7cm, width=6cm, angle=-90]{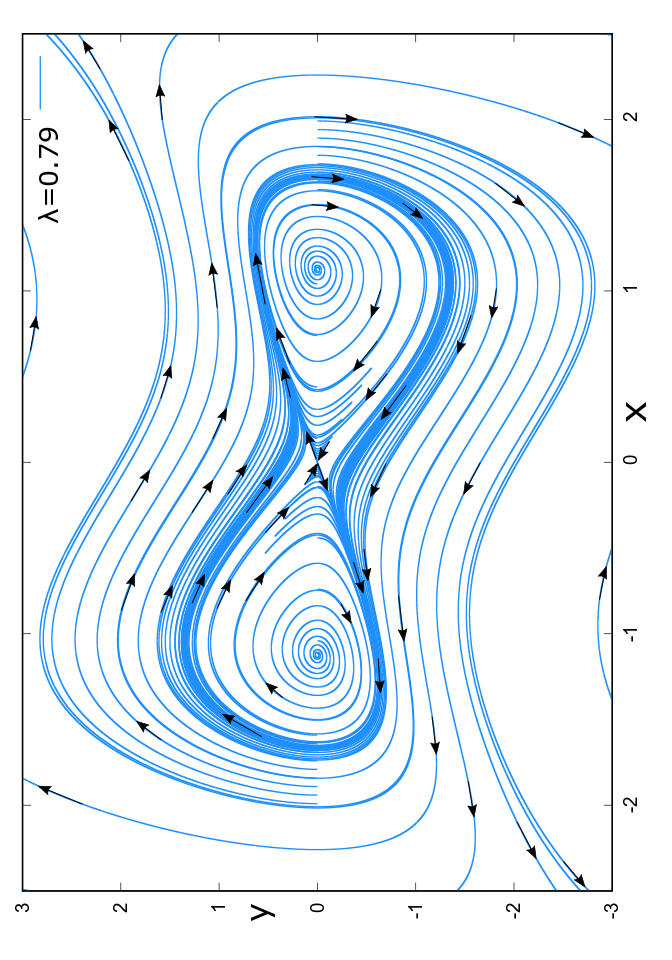}
\caption{$\lambda=0.79,~k=-1$,  the stable limit cycle and the unstable limit cycle collide with each other.}
\end{figure}

\begin{figure}[hbtp]
%\centering
\includegraphics[height=8.7cm, width=6cm, angle=-90]{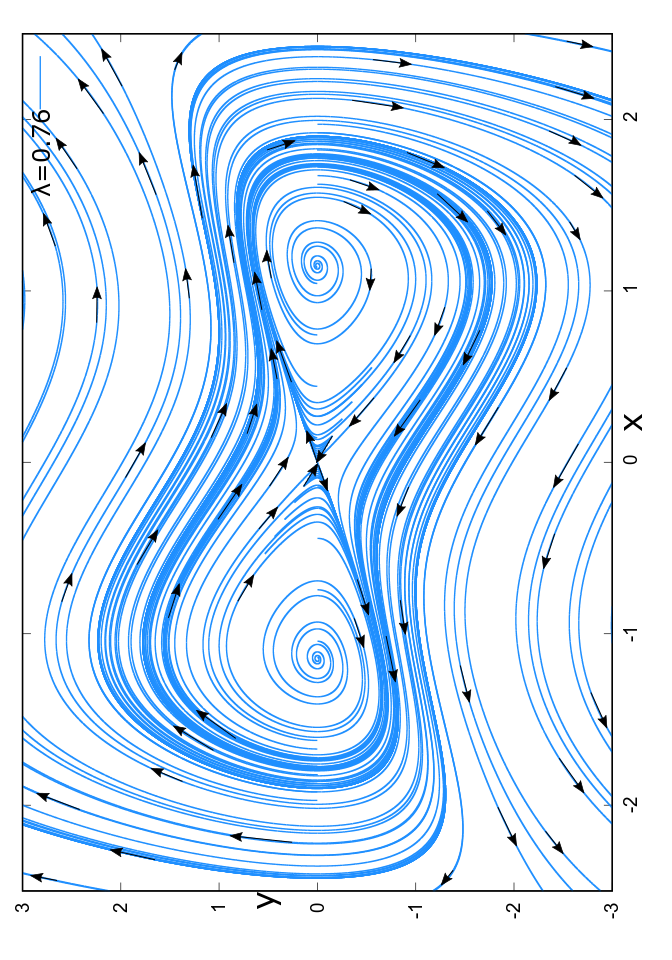}
\caption{$\lambda<0.79,~k=-1$, no limit cycle in the phase portrait, the limit cycles disappear.}
\end{figure}
As $\lambda$ approaches $1$, Hopf bifurcation occurs; and as $\lambda$ goes just below $1$, two new limit cycles of same kind occurs around the wells and these two limit cycles are of opposite kind to the boundary limit cycle. Hence at $\lambda\lesssim 1$, there are three limit cycles in the phase portrait of the system.
At $\lambda\lesssim 1$, the growth of the amplitude of the limit cycles surrounding the wells is given by equation (9). The bigger boundary limit cycle of opposite kind to these two limit cycles shrink as $\lambda$ becomes smaller. As $\lambda$ decreases, these two limit cycles around the two wells eventually collide with each other at the saddle point $(0,0)$ forming a homoclinic orbit. For $k=-1$, this event occurs at $\lambda\approx 0.82$. The time period of the orbit approaches infinity. As $\lambda$ becomes smaller, we find numerically that these two limit cycles merge together and form a bigger limit cycle of same kind.
At this point, there are two limit cycles of opposite kind in the phase portrait. Now as $\lambda$ becomes smaller, this bigger limit cycle which is formed due to the merging of the two limit cycles (previously formed by Hopf bifurcation around the wells) grow bigger and the other one, surrounding it, of the opposite kind which has been present for all the values of $\lambda$ up to this point, becomes smaller. Eventually, these two limit cycles of opposite kind collide with other. We find  that for $k=-1$, this event happens at $\lambda\approx 0.79$. As $\lambda$ becomes just smaller than this value two limit cycles are being destroyed. At this point, there is no limit cycle in the phase portrait of the system, there are only fixed points. The phase space trajectories are denser at the region where the destruction of two limit cycles occur. The signature of the stable limit cycle which was there for greater value of $\lambda$ is still present there even in it's absence. Figure 6 clearly demonstrates that.
\begin{figure}[hbtp]
%\centering
\includegraphics[height=8.5cm, width=6cm, angle=-90]{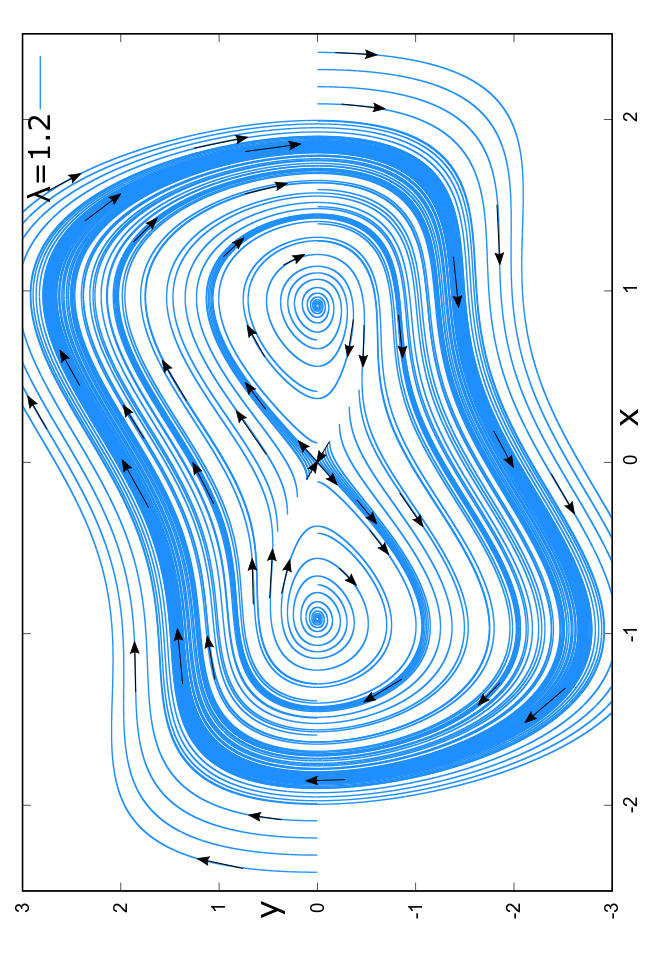} 
\caption{$\lambda=1.2,~k=1$; one stable limit cycle and at $\lambda=1$, Hopf bifurcation occurs.}
\end{figure}
For positive $k$, $k=1$; phase portrait for only two values of $\lambda$ is showed for brevity. For $k=1$ and $k=-1$, the incidents of creation of homoclinic orbit, merging of the two newly formed limit cycles due to Hopf bifurcation, the destruction of the limit cycles would happen at the same values of $\lambda$ due to symmetry.
\begin{figure}[hbtp]
%\centering
\includegraphics[height=8.7cm, width=6cm, angle=-90]{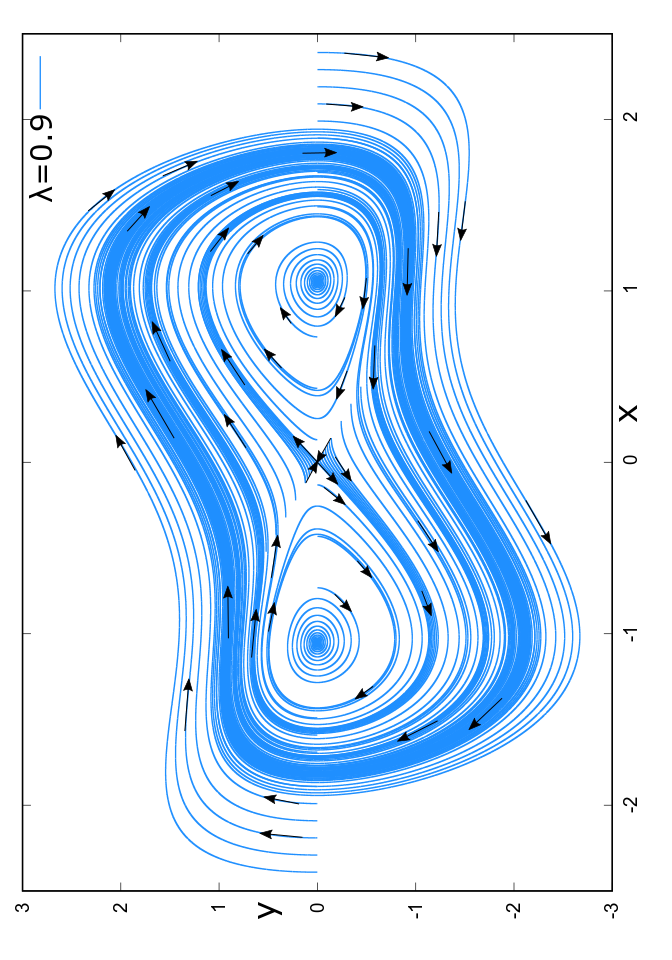} 
\caption{$\lambda=0.9,~k=1$; one stable limit cycle and two newly formed unstable limit cycles.}
\end{figure}
The distance between the two spiral type fixed points (for now, we are assuming $k$ to be small, large $k$ behaviour is being discussed in details in the next section) increases as $\sim \frac{1}{\sqrt{\lambda}}$  with decreases in $\lambda$ for both positive and negative values of $k$. One thing to notice by comparing figure 7 and figure 8 with the figures 1-6 is that, the overall orientation of the phase curves is not same. This is due to the following reason. From equation (2), on $y-$axis,
\begin{align*}
& \dot{x}=y\\
& \dot{y}=ky\\
& \Rightarrow \left(\frac{dy}{dx}\right)_{x=0}=k
\end{align*}
Therefore, for positive $k$, on $y-$axis, the slope of the phase curves are positive and for negative $k$, on $y-$axis the slope of the phase curves are negative. Most coincidentally, the slope of all the phase curves in the phase portrait on $y-$ axis are same for a fixed value of $k$. \\
%When one looks at the events described from figure 1 to figure 6, i.e; with the decreasing order in $\lambda$, the %bifurcations seem quite natural but if one looks at the events with increasing $\lambda$, i.e; from figure 6 to figure %1, there is an interesting thing happening at $\lambda=0.79$, two limit cycles of opposite type is being created. %There must be something going on with the fixed points. We have suggested a reason behind this in the section 3. 
\subsection*{Large $k$ behaviour}
At very large $k$, one can calculate the time period of the boundary limit cycle. From equation (1), one can write
\begin{align}
&\frac{d}{dt}\left[\dot{x}+k\left(\frac{x^3}{3}-x\right)\right]+(\lambda x^3-x)=0\nonumber\\
&\dot{x}=k\left[z-\left(\frac{x^3}{3}-x\right)\right]\\
&\dot{z}=-\frac{1}{k}(\lambda x^3-x)\\
&{\rm where~} z=\frac{1}{k}\dot{x}+\left(\frac{x^3}{3}-x\right)\nonumber
\end{align}
Let's for now, assume $k>0$ and $k>>1$.
From equation (12) and equation (13), $\dot{z}$ is much smaller compared to $\dot{x}$, until and unless $z$ becomes nearly equal to $\left(\frac{x^3}{3}-x\right)$ in the phase plane of $z-x$. There are fast trajectories parallel to $x-$axis in this region depicted in figure 9. As any phase curve tends to the curve $\left(\frac{x^3}{3}-x\right)$, $\dot{z}$ and $\dot{x}$ stats becoming comparable and as eventually any phase curve touches the curve, $\left(\frac{x^3}{3}-x\right)$, $\dot{x}$ becomes zero. $\dot{z}$ may or may not have a small finite value depending on the point of touching. For a small finite value of $\dot{z}$, equation (12) and equation (13) suggests that a slow phase space motion will be occurring along the curve $\left(\frac{x^3}{3}-x\right)$ for $x>1$ and $x<-1$.   
\begin{figure}[hbtp]
\includegraphics[width=8.4cm, height=7.5cm]{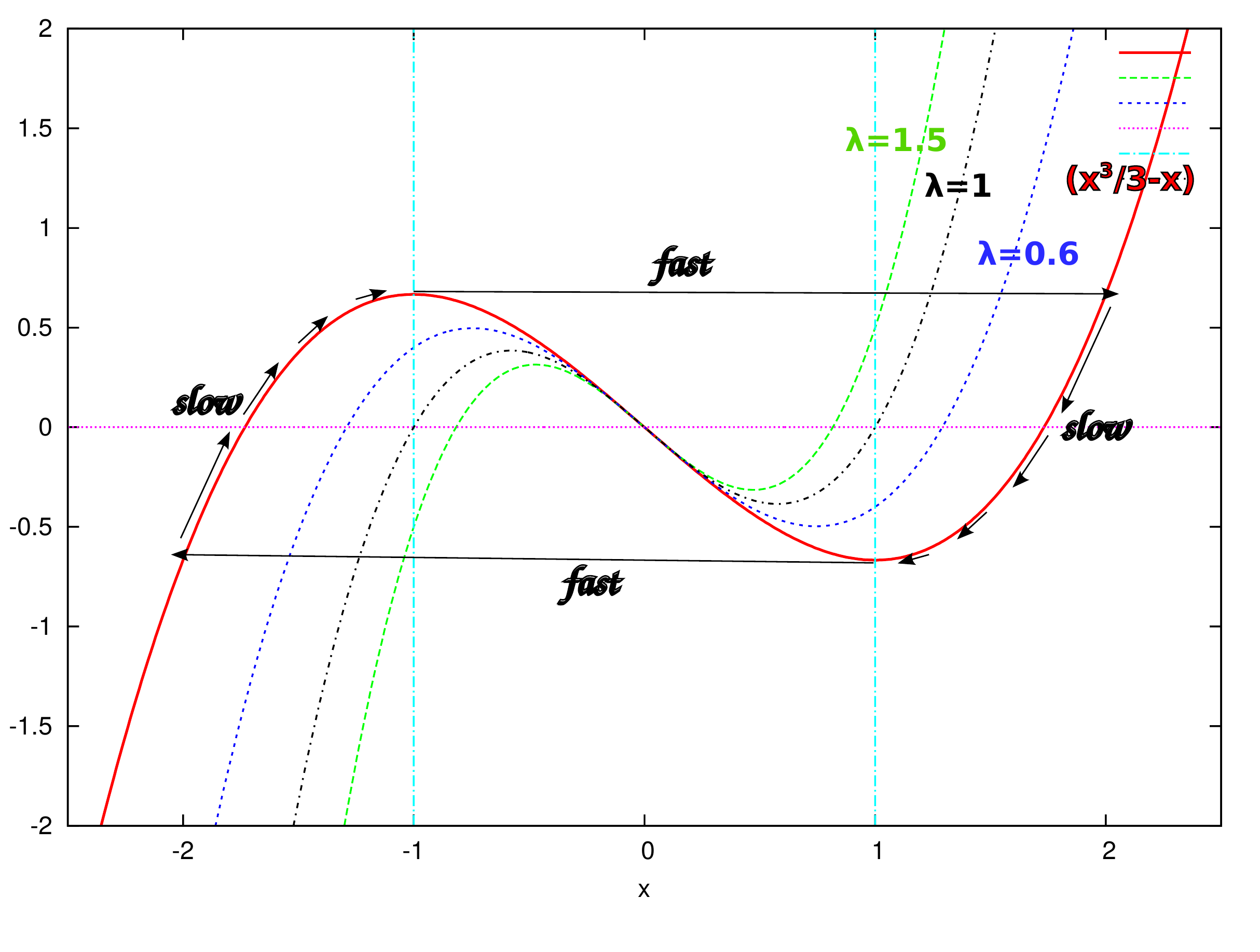}
\caption{slow-fast oscillation, possibility of a stable periodic orbit for $\lambda>1,~k>0$.}
\end{figure} 
If any trajectory manages to reach any of the extremal points ($x=\pm 1$) of the curve $\left(\frac{x^3}{3}-x\right)$ such that at this point $\dot{x}>0$ for $z>0$ or $\dot{x}<0$ for $z<0$, the tangent motion along that extrema (at $x=\pm 1$) will eventually draw the trajectory in the fast region, and the trajectory will hit again another slow-motion arm of $\left(\frac{x^3}{3}-x\right)$ undergoing a slow motion again and thus a periodic orbit is formed. From equation (12), $\dot{x}=0$ corresponds to $z=\left(\frac{x^3}{3}-x\right)$ curve and  $\dot{z}=0$ corresponds to $x=0,~\pm\frac{1}{\sqrt{\lambda}}$ straight lines. The fixed points are on the  $z=\left(\frac{x^3}{3}-x\right)$ curve, i.e; at $(0,0),~\left(\frac{1}{\sqrt{\lambda}},\pm\frac{1}{\sqrt{\lambda}}(\frac{1}{3\lambda}-1)\right)$.  Now for $\lambda<1$, as the two fixed points on the curve $\left(\frac{x^3}{3}-x\right)$ are at the right side of $x=1$ and on the left side of $x=-1$ respectively. Any phase space trajectory eventually settles down on any of the two fixed points of the slow-motion arms and thus prohibiting the formation of the stable limit cycle. Hence the condition of getting a periodic orbit is as follows,
\begin{align}
\frac{1}{\sqrt{\lambda}}\leqslant 1\nonumber\\
\Rightarrow\lambda\geqslant1
\end{align} 
Therefore, for very large value of $k$, the boundary limit cycle exists only for $\lambda\geqslant 1$. The same conclusion can be drawn in the same fashion by also considering negative $k$.\\
For $\lambda>1$, the main contribution in the time period comes from the motion on the slow-arms. On the slow arms $z\sim\left(\frac{x^3}{3}-x\right)\Rightarrow\dot{z}=(x^2-1)\dot{x}$. Now using equation (13), we find the time period $T$.
\begin{eqnarray}
& &T\simeq2k\int_1^2dx\frac{(x^2-1)}{(\lambda x^3-x)}\nonumber
\end{eqnarray}
\begin{align}
&=\frac{2k}{3\lambda}(3\lambda-1)ln2+\frac{2k}{3\lambda}ln\left[\frac{2(\lambda-4)}{(\lambda-1)}\right]\nonumber\\
&+\frac{k}{3\lambda}ln\left[\frac{(2\sqrt{\lambda}-1)}{(\sqrt{\lambda}-1)}\right]+\frac{k}{3\lambda}ln\left[\frac{(2\sqrt{\lambda}+1)}{(\sqrt{\lambda}+1)}\right]
\end{align}
The $2$ factor is coming in front of the integration because there are two slow arms. The lower limit in the integral, i.e; $x=1$, corresponds to the positive $x$ position of the minima of the function $\left(\frac{x^3}{3}-x\right)$ and the upper limit of the integral is $x=2$ because $x=2$ is the rightmost extent of the limit cycle which is found by drawing tangent on the maxima of the curve, $\left(\frac{x^3}{3}-x\right)$ and then finding the intersection of that tangent with the slow arm in positive $x$ plane. Let's define a percentage of error in finding time period of the orbit.
\begin{equation}
n=\frac{(T'-T)}{T}\times 100\%
\end{equation}
where $T'$ is the time period of the orbit computed numerically. $k=10,~n=59.41\%;~k=100,~n=25.18\%;~k=200,~n=24.02\%$.
\subsection*{Observations}
If one looks at the figures from 1 to 6 in reverse order, i.e; in the increasing order of $\lambda$, there must be something going on near $\lambda=0.79$ because in the phase portrait two limit cycles of opposite kind appear. With the increase in $\lambda$, one of the limit cycle grows in size and the other one diminishes at $\lambda=1$ by Hopf bifurcation. We have given an explanation about the creation of two limit cycles of opposite kind in the phase portrait in the next section.\\
Let's say, at $\lambda=\lambda_1$, Hopf bifurcation occurs; at $\lambda=\lambda_2$, the limit cycles, created due to Hopf bifurcation, collide with each other at the saddle point, homoclinic orbit is formed; at $\lambda=\lambda_3$, the limit cycles in the phase portrait are destroyed. The value of $\lambda_1$ is always 1, i.e; it does not depend on $k$, which is seen from the matrix equation (4). We numerically find that for $k=\pm 1,~\lambda_2\simeq 0.82, ~\lambda_3\simeq 0.79$; for $k=\pm 0.1,~\lambda_2\simeq 0.80085, ~\lambda_3\simeq 0.76$ and for $k=\pm 2,~\lambda_2\simeq 0.8665, ~\lambda_3\simeq 0.851$. Hence as the value of $k$ goes higher and higher, $\lambda_2,~\lambda_3$ are shifted towards $\lambda_1$, i.e, 1. We've given a qualitative explanation about this phenomena in the following section.
\section{The Bifurcations with the variation in $k$ and $\lambda$}
From equation (3), we see that the local phase portrait near $(0,~0)$, is always saddle for any real values of $k$ and $\lambda$. This is not the case for the fixed points $(\pm\frac{1}{\sqrt{\lambda}},~0)$. The eigen values of matrix $\hat{B}$ are
\begin{align}
\alpha_{+,-}=\frac{1}{2}\left(Tr(\hat{B})\pm\sqrt{Tr(\hat{B})^2-4\det(\hat{B})}\right)
\end{align}
where $Tr(\hat{B})=k\left(1-\frac{1}{\lambda}\right)$ and $det(\hat{B})=2$.
Let's consider the case $k>0$. Now the term in the square root may be positive, negative or zero and the local phase portrait of the fixed point will be accordingly node, spiral or star node. Initially we assumed $k$ to be small and hence the fixed points $(\pm\frac{1}{\sqrt{\lambda}},~0)$ were spiral. Now, we find the conditions to get node.
\begin{equation}
Tr(\hat{B})^2-4\det(\hat{B})\geqslant 0
\end{equation}
\begin{align}
&\Rightarrow k\geqslant\frac{2\sqrt{2}}{1-\frac{1}{\lambda}}~~~~{\rm for \lambda>1}\\
&{\rm and~} k\geqslant-\frac{2\sqrt{2}}{1-\frac{1}{\lambda}}~~~~{\rm for \lambda<1}
\end{align}
\begin{figure}[hbtp]
\includegraphics[scale=0.375]{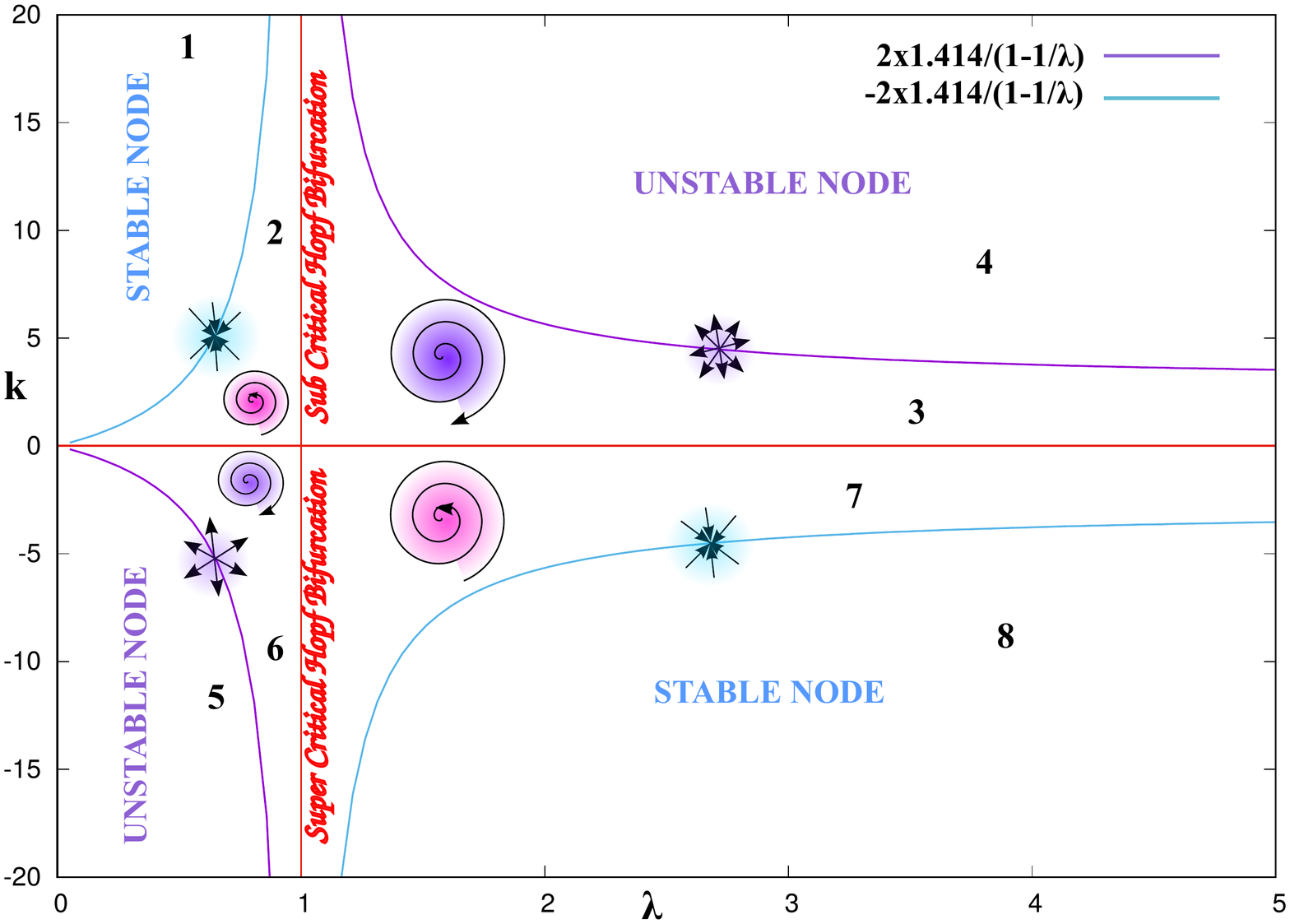} 
\caption{Local phase space behaviour near the fixed points $(\pm\frac{1}{\sqrt{\lambda}},~0)$. }
\end{figure}
In the $k-\lambda$ parameter space, region 1, in the figure 10, corresponds to the region of stable node; region 2 corresponds to stable spiral and the boundary between region 1 and the region 2 corresponds to stable star. $\lambda=1$ line corresponds to the Hopf bifurcation. Region 4 in the phase portrait corresponds to unstable node; region 3 corresponds to unstable spiral and the border line of region 3 and region 4 corresponds to unstable star.
For $k<0$, similar arguments give the condition to get node.
\begin{align}
&\Rightarrow k\leqslant-\frac{2\sqrt{2}}{1-\frac{1}{\lambda}}~~~~{\rm for \lambda>1}\\
&{\rm and~} k\leqslant\frac{2\sqrt{2}}{1-\frac{1}{\lambda}}~~~~{\rm for \lambda<1}
\end{align}
From figure 10, one can see that each of the the fixed points $(\pm\frac{1}{\sqrt{\lambda}},~0)$ under go an another change below $\lambda=1$, becomes fixed point of node type to a fixed point of spiral type. Greater the value of $k$, closer the value of $\lambda$ to 1, for such a change. Qualitatively, for greater $\lambda$, both of the fixed points $(\pm\frac{1}{\sqrt{\lambda}},~0)$ become spiral to node much quickly. The bifurcations are much more drastic for higher values of $k$. As $k\rightarrow\infty$, this change happen at $\lambda\rightarrow 1$. This fact qualitatively explains both of the observations in the previous section. This fact also explains impossibility of getting no limit cycle in the phase portrait for $\lambda<1$ as $k$ goes to infinity. This change of spiral to node for $\lambda<1$, occurs at $\lambda\simeq0.26$ for $k=\pm1$; at $\lambda\simeq0.034$ for $k=\pm0.1$ and at $\lambda\simeq0.41$ for $k=\pm2$; calculated from the equality of equation (20) and equation (22). $\lambda_3$s for these values of $k$ are greater than these values.
\section{Asymmetric double well Van der Pol Oscillator}
Equation of motion of a Van der Pol oscillator in an asymmetric double well can be written as
\begin{equation}
\ddot{x}+k\dot{x}(x^2-1)+\lambda x\left(x-\frac{a}{\sqrt{\lambda}}\right)\left(x+\frac{b}{\sqrt{\lambda}}\right)=0
\end{equation}
where $a,~b$ are positive real number. $a=b$ gives symmetric double well.
\begin{figure}[hbtp]
\includegraphics[scale=0.34, angle=-90]{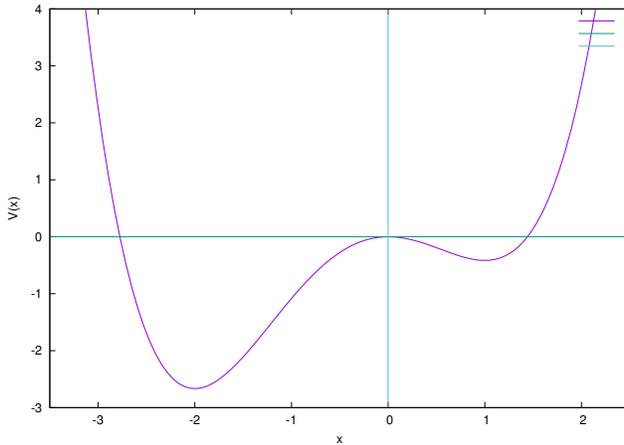} 
\caption{Asymmetric double well with $a=1$ and $b=2$.}
\end{figure}
We discuss briefly this case because the analysis is more or less same as the previous one. 
Unlike the previous case of symmetric double well, in this case, Hopf bifurcation occur at two  different values of $\lambda$. At $\lambda=a^2$, fixed point $(\frac{a}{\sqrt{\lambda}},0)$ to one of the wells on the positive half plane of $x$, undergoes a Hopf bifurcation. $\lambda\lesssim a^2$, a limit cycle appears around this fixed point. Just like the previous case stability of the fixed point, stability of this limit cycles depend on the sign of $k$. Similarly at $\lambda=b^2$, fixed point $(-\frac{b}{\sqrt{\lambda}},0)$ corresponding to the other well on the negative half plane of $x$, undergoes a Hopf bifurcation. $\lambda\lesssim b^2$, another limit cycle appears around this fixed point. One can calculate amplitude and frequency of the newly born limit cycles for small $k$ in a similar fashion as before. The limit cycle motion around the fixed point $(\frac{a}{\sqrt{\lambda}},0)$ can be written as 
\begin{equation}
x(t)\simeq\frac{a}{\sqrt{\lambda}}+A_a~cos\left(\omega'_at+\Phi_a\right)
\end{equation}
where
\begin{align}
& A_a=2\sqrt{\left(\frac{a^2}{\lambda}-1\right)}\\
&\omega'_a=\left(\omega_a+\frac{3}{2\omega_a}\left(\frac{a^2}{\lambda}-1\right)\right)\\
&\omega_a=\sqrt{a(a+b)}\\
&{\rm and~} \Phi_a {\rm ~is~a~constant~phase~factor.}\nonumber
\end{align}
$\omega_a$ is the natural frequency of small oscillations around $x=\frac{a}{\sqrt{\lambda}}$ in the absence of dissipative term in the equation of motion.
As this approximation method to find the values of amplitude and frequency of the limit cycle holds good at $\lambda\lesssim a^2$, we write $\lambda=a^2-\epsilon$. Hence
\begin{eqnarray}
A_a=\frac{2}{a}\sqrt{\epsilon}\\
\omega'_a=\omega_a+\frac{3\epsilon}{2a^2\omega_a}
\end{eqnarray}
We find similar expressions for the Hopf bifurcation at $\lambda=b^2$. The expressions for amplitude and frequency, very similar to the expressions (25) to (29), can be found by replacing $a$ by $b$ and $b$ by $a$ in the equations (25) to (29). Let's define a two element set $\Lambda$ to be $\{a^2,b^2\}$. The supremum of the set $\Lambda$ is $sup~\Lambda$, the infimum of the set $\Lambda$ is $inf~\Lambda$. At $\lambda\lesssim sup~\Lambda$, amplitude and frequency of the limit cycle created due to Hopf bifurcation at $\lambda=sup~\Lambda$; are $sup~A$ and $sup~\omega$ respectively. Similarly amplitude and frequency of the limit cycle created due to Hopf bifurcation at $\lambda=inf~\Lambda$ are $inf~A$ and $inf~\omega$. From equation (28), we see that $sup~A$ grows slower than $inf~A$. The physical explanation of this fact is that deeper the well, steeper the well is and hence slower the growth of the limit cycle, around the fixed point corresponding to that well, is. From equation (27), we find that the natural frequency part in $sup~\omega$ is bigger than the natural frequency part in $inf~\omega$ but the correction terms of the natural frequency is the opposite story.\\ 
Now the bifurcation events for finite value of $k$ can be stated as follows: for $\lambda>sup~\Lambda$, there exists only one limit cycle in the phase portrait, we call it by boundary limit cycle. The stability of this limit cycle depends on the sign of $k$ as before. As $\lambda$ decreases to the value $sup~\Lambda$, Hopf bifurcation occurs, the stability of the fixed point corresponding to that well is changed. For $inf~\Lambda<\lambda<sup~\Lambda$, there exists two limit cycles of opposite kind in the phase portrait, one is the boundary limit cycle and the other one is the limit cycle inside it around the corresponding fixed point. As $\lambda$ approaches $inf~\Lambda$, second Hopf bifurcation occurs. For $\lambda_2<\lambda<inf~\Lambda$, there are three limit cycles in the phase portrait, one is the bigger boundary limit cycle and the other two, opposite kind to the boundary one, are the two limit cycles created by two Hopf bifurcations. The limit cycles created due to Hopf bifurcations grow in size with the decrease in $\lambda$ and the boundary limit cycle shrinks towards these two limit cycles as $\lambda$ decreases. At $\lambda=\lambda_2$, the two limit cycles of same kind, formed due to Hopf bifurcations, collide with other at the saddle point making a homoclinic orbit. As $\lambda$ goes smaller one single limit cycles is formed out of this two limit cycles of same kind. For $\lambda_3<\lambda<\lambda_2$, there are two limit cycles in the phase portrait as before. At $\lambda=\lambda_3$, the limit cycle formed by merging of two limit cycles collide with the boundary limit cycle of opposite kind. There is no limit cycle in the phase portrait for $\lambda<\lambda_3$.\\
Similarly, one can explore large $k$ behaviour, we find by similar arguments presented previously that for very large $k$ limit, limit cycle exists only for $\lambda>sup~\Lambda$. For $\lambda>sup~\Lambda$, we calculate the time period of the boundary limit cycle.
\begin{eqnarray}
& & T\simeq\frac{2k}{\lambda}\left[\frac{(b^2-\lambda)}{b(a+b)}ln\left(\frac{2\sqrt{\lambda}+b}{\sqrt{\lambda}+b}\right)\right]\nonumber\\
& & +\frac{2k}{\lambda}\left[\frac{(a^2-\lambda)}{a(a+b)}ln\left(\frac{2\sqrt{\lambda}-a}{\sqrt{\lambda}-a}\right)\right]\nonumber\\
& & +\frac{2k}{ab}ln2
\end{eqnarray}
\section{Summary and Conclusion}
We find that the shape factor $\lambda$ governs the behaviour of the phase curves in the phase portrait. Qualitative behaviour of the phase portrait very much depends on the local phase space behaviour near the two fixed points corresponding to the wells of a double well. If one goes at higher values of $k$, phase portrait changes with $\lambda$ within shorter interval of $\lambda$. The bifurcations become more abrupt. \\
We can extend this work by considering potentials with more wells. From our current understandings and experiences about double well Van der Pol oscillator, the bifurcation events would be qualitatively similar, i.e; there would be more limit cycles created due to Hopf bifurcations in different wells. Again, one can consider the case of asymmetric double well with $a$ or $b$ to be zero. We anticipate from our current understanding that the events qualitatively wouldn't differ much, i.e; in that case we would have to work with the Hopf bifurcation and the stability of one well present in the system. 
\section{Acknowledgement}  
The author is thankful to Prof. Jayanta Kumar Bhattacharjee for useful discussions and suggestions.

\end{document}